\documentclass[12pt, preprint]{emulateapj}

\shortauthors{Shi et al.}

\begin{document}

\title{ Thermal and Non-Thermal Infrared Emission from M87 }

\author{Y. Shi\altaffilmark{1}, G. H. Rieke\altaffilmark{1}, D. C. Hines\altaffilmark{2}, 
K. D. Gordon\altaffilmark{1}, E. Egami\altaffilmark{1}} 

\altaffiltext{1}{Steward Observatory, University of Arizona, 933 N Cherry Ave, Tucson, AZ 85721, USA}
\altaffiltext{2}{Space Science Institute 4750 Walnut Street, Suite 205,
Boulder, Colorado 80301}

\begin{abstract}
We discuss images of M87 from 3.6 to 160 $\mu$m obtained with {\it Spitzer}. 
As found previously, there is an excess in the far infrared over a
simple power law interpolation from the radio to the resolved nonthermal
features in the mid-infrared and optical. We show that this excess
is most likely warm dust in the galaxy itself, and that the properties
of this emission component are similar to the far infrared emission
of normal giant elliptical galaxies. The new observations let us
determine the spectrum of the jet and surrounding lobes of nonthermal
emission. We find that even in the lobes the synchrotron break frequency
is in the optical, probably requiring {\it in situ}
particle acceleration not only in the jet but in the lobes as well.

\end{abstract}


\section{Introduction} 

Fifty years  ago, Baade (1956)  discovered that the  highly collimated
jet  in  M87  (NGC  4486,  3C  274,  Virgo  A)  glows  in  synchrotron
emission. Since  then, we  have learned that  such jets are  often the
most spectacular manifestation of an active galactic nucleus (AGN) and
that the resulting  beaming of energy strongly influences  our view of
the   entire  AGN   phenomenon.  Because   it  is   relatively  nearby
\citep[$d$=16Mpc,  78 pc  arcsec$^{-1}$;][]{Tonry91}, the  jet  in M87
remains one of the most  readily studied of these structures.  Much of
our understanding of jets rests on the high-resolution, deep images of
it now available from the radio through the X-ray.

In  the  radio, an  inner  region (${1\farcm5}{\times}{2\farcm0}$)  is
embedded  in   a  halo  with   a  diameter  up   to  $\sim$$15\farcm0$
\citep[e.g.][]{Mills52, Baade54, Rottmann96}.  This region is composed
of  the nucleus,  jet, North-preceding  (Np) lobe  and South-following
(Sf) lobe \citep{Turland75,  Forster80}.  High resolution radio images
of  the inner  region  have revealed  complex  structure, with  knots,
filaments, edges  and rings \citep[e.g.][]{Owen80,  Hines89}.  A radio
outflow originating from the nucleus terminates at the boundary of the
radio  halo  \citep{Owen00}.   In   the  optical,  {\it  Hubble  Space
Telescope} ({\it HST}) images reveal apparent superluminal motions for
several knots in the  jet \citep{Biretta99}.  The images also indicate
differences  between  the  morphologies   of  the  optical  and  radio
synchrotron  emission \citep{Sparks96, Perlman01a}.   M87 sits  at the
center   of  the   hot   X-ray  atmosphere   of   the  Virgo   cluster
\citep{Fabian94},  with  the  gas  temperature decreasing  toward  the
central  region.    However,  simple  cooling  flow   models  are  not
consistent  with recent X-ray  observations, such  as the  low cooling
rate   and  the   single   temperature  phase   of  intracluster   gas
\citep[e.g.]{Bohringe01,   Molendi01}.   In   the   subarcsecond  {\it
Chandra} X-ray  image, two spectacular  outflow-like structures extend
from the nucleus east and southwest, respectively, possibly created by
buoyant bubbles \citep[e.g.][]{Young02, Forman06}.

Ground-based mid-IR observations have  only detected the emission from
the nucleus  and the  brightest knots of  the jet, and  are consistent
with synchrotron  emission models \citep{Perlman01b,  Whysong04}.  The
{\it  Infrared  Space  Observatory}  ({\it ISO})  mid-IR  images  have
marginally detected the  lobe \citep{Xilouris04}. Other features, such
as the individual  filaments in the lobe, are  only detected weakly in
the  near-IR, but  not  yet in  the  mid-IR \citep{Stocke81,  Smith83,
Neumann95, Perlman01a}.  {\it ISO}  and the {\it Infrared Astronomical
Satellite} ({\it IRAS}) observations of M87 reveal excess far-infrared
emission   above  the   synchrotron  interpolation   from   the  radio
\citep{Haas04, Xilouris04}.  However, the available information in the
infrared currently lags far behind  the exquisite and very deep images
available  in the  radio, optical,  ultraviolet, and  X-ray.   To help
remedy this situation, we present  infrared images of M87 at 3.6, 4.5,
5.8, 8,  24, and 70$\mu$m and  photometry at 160  $\mu$m obtained with
the Infrared  Array Camera \citep[IRAC;][]{Fazio04}  and the Multiband
Imaging Photometer \citep[MIPS;][]{Rieke04} on {\it Spitzer}.

\section{Data Reduction and Analysis}
\subsection{HST image}

We retrieved archived  {\it HST} data (PID 8048,  PI J. Biretta) taken
with the  Wide Field  and Planetary  Camera  2 (WFPC2).   The images  were
obtained on 2001  November 17 through filters F450W,  F606W and F814W.
The data  were processed through the Post  Observation Data Processing
System  (PODPS)  pipeline  to  remove bias  and  flat-field  artifacts
\citep{Biretta96}.   Individual  exposures  at  each  wavelength  were
combined  to remove  the cosmic-ray  events. The  outputs of  the four
chips of  WFPC2 were combined and  the final image was  rotated to put
north at the top and east to the left.

To obtain  the optical  image of  the jet and  lobe, galaxy  light was
subtracted  in  IRAF as  in  \citet{Perlman01a}.   We  masked out  the
optical jet and the Sf lobe as indicated by the 5 GHz radio image.  We
also masked the region without data  because of the smaller FOV of the
PC chip compared to the WFC one.  The task ELLIPSE in IRAF was used to
fit isophotes  of the galaxy image  and a model  image was constructed
with  task BMODEL.  The  optical nonthermal  emission was  obtained by
subtracting the model image from  the original image with task IMCALC.
We then masked out the globular clusters and chip joining region based
on  the residual  image.  The  above processes  were  repeated several
times  until  the  final   galaxy-subtracted  image  did  not  exhibit
ring-like  structure.   In   this final image,  we
replaced each  pixel of the masked  region except for  the jet
and lobe with a simulated value based on the average and scatter of
residual counts in the non-masked region with the same distance to the
galactic center.   The final  uncertainty includes the  fluctuation in
the residual  image, flat-fielding errors ($\sim$1\%)  and zero-point errors
($\sim$1\%).

\begin{figure}
\epsscale{1.}
\plotone{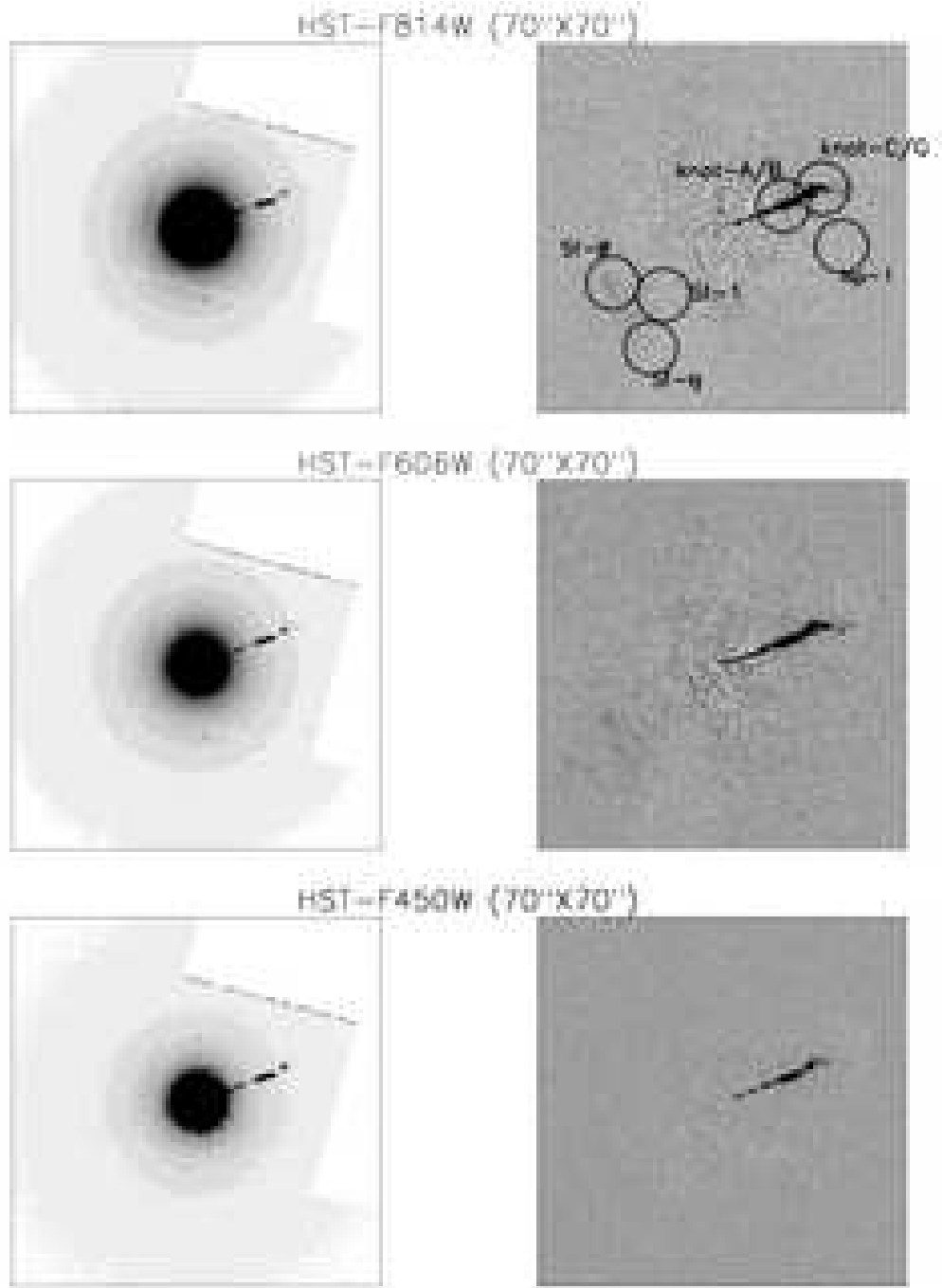} 
\caption{\label{HST_All_Bands}  
Original {\it HST}  images of M87 in the  left column and galaxy-light
subtracted image in the right column.
}
\end{figure}

As  shown  in  the  right column  of  Figure~\ref{HST_All_Bands},  the
optical jet is well detected  at all {\it HST} bands after subtracting
the galaxy  light.  In the  region of the  radio Sf lobe, there  is an
elongated  feature   that  becomes   more  prominent  at   the  longer
wavelengths.  This feature has morphology, location and position angle
similar  to  the radio  component  $\theta$ seen  in  the  6 cm  image
\citep[See][]{Hines89}.   The  optical spectrum  of  this feature  has
$\alpha$ $\sim$  2 (where $f_\nu ~ {\propto} ~ {\nu}^{-\alpha}$), steeper
than that of the optical jet \citep[$\alpha \sim$1, See][]{Perlman01a}
but flatter  than that of  galaxy light ($\sim$ 2.5).  The non-thermal
spectrum of  this feature  indicates it is the optical component of 
the Sf  lobe.  The optical counterpart of filament $\theta$ has
also   been    detected   by   previous    ground-based   observations
\citep{Stiavelli92, Sparks92}.

\subsection{{\it SPITZER} DATA}

Our  MIPS observations  (PID  82, PI  G.   Rieke) were  made with  the
standard small  field photometry mode  and were reduced with  the MIPS
instrument    team   Data   Analysis    Tool   (DAT)    version   3.03
\citep{Gordon05}. At 70 $\mu$m, extra processing steps beyond those in
the DAT were applied to remove known transient effects associated with
the   70    $\mu$m   detectors   to   achieve    the   best   possible
sensitivity. Detector-dependent structures were removed by subtracting
column averages from  each exposure with the source  region masked. In
addition, a  pixel-dependent time filter  was applied (with  the source
region masked)  to remove small pixel-dependent  residuals.  
For all the bands, the  background was subtracted  using the sky  level at
the edge of each field.  The resultant images were resampled by a factor
of four and rotated to put north at the top and east to the left. The final
MIPS     images     have      fields     of     view     (FOVs)     of
${7\farcm5}{\times}{7\farcm5}$,         ${6\farcm5}{\times}{3\farcm0}$ and
${2\farcm5}{\times}{6\farcm0}$ at 24 $\mu$m, 70 $\mu$m and 160 $\mu$m,
respectively.  

To isolate  the output  of the  jet and lobes at 24 and 70$\mu$m, we  subtracted the
nuclear emission.   We used the empirical point  spread function (PSF)
based on the observed images of stars and provided by the {\it Spitzer}
Science
Center (SSC) \footnote{http://ssc.spitzer.caltech.edu/mips/psf.html}.      We
aligned  the PSF with  the centroid  of the  image and  determined the
normalization  by matching  the  average surface  brightness within  a
half-FWHM radius.  Although  the PSF image has a  FOV smaller than the
image of M87,  the contribution is negligible compared  to the Poisson
noise at regions outside  of the PSF coverage.  The original and nuclear-subtracted
images are shown in Figure~\ref{MIPS}. The uncertainty in the nuclear-subtracted image was
estimated  by repeating  the subtraction  with the  PSF placed  over a
4$\times$4 grid  around the centroid  of the image.   This uncertainty
and the  Poisson noise  were added quadratically  to create  the final
noise image.

IRAC  data were taken  from the  {\it Spitzer}  archive (PID  3228, PI
W. Forman).  We used the  post-basic-calibrated-data (post-BCD) images
at a pixel  scale of 1$\farcs$2 pixel$^{-1}$. The  resultant images were
also rotated to put north at the top and east to the left.

To obtain  the non-thermal emission of  the jet and lobes  at the IRAC
bands, the stellar emission was  subtracted similarly to the {\it HST}
image. Figure~\ref{IRAC}  shows the original image  and residual image
of   M87  at   the  four   IRAC  bands.    For  some   of   the  faint
low-surface-brightness  features, we  found it  impossible  to extract
reliable measurements  at 3.6 and  4.5$\mu$m because of  the residuals
from  subtraction of the  stellar emission.   With this  exception, we
obtained  reliable measurements for  two positions  along the  jet and
four in  the lobes at all the  {\it Spitzer} bands out  to 24$\mu$m, a
measurement  of the jet  and Sf  lobe at  70$\mu$m, and  an integrated
measurement of the nuclear region and surrounding galaxy at 160$\mu$m.

Because of the  complexity of the images, we  could not apply standard
aperture corrections to our photometry.  Instead, we used a variety of
approaches optimized for each  situation.   The  nuclear fluxes
were  measured by PSF  fitting and  calibrated against  standard stars
measured the same way.  To determine spectral slopes, we convolved all
the images to  the resolution of the lowest  resolution image and then
used the  same photometry approach  on all.  To obtain  flux densities
for the  jet knots, we used the  {\it HST} F814W image  as a template.
We   first  did   photometry   in  the   selected   aperture  on   the
full-resolution image,  and then repeated the photometry  on the image
convolved with the 24$\mu$m PSF.   A comparison of the results yielded
the appropriate aperture correction.  We quote all the IRAC photometry
at this resolution also for consistency. However, the nonthermal lobes
are not well  enough detected to use the F814W image  in this way, and
they are  more extended  than the knots,  so we report  photometry for
them with  no aperture correction.  For IRAC measurements of  the full
M87 galaxy, we applied  extended source corrections in accordance with
the procedure  recommended by the SSC. MIPS
photometry  of  the whole  galaxy  was  corrected  in accordance  with
smoothed STinyTim  models of the image,  which have been  shown to fit
the observed PSFs well (Engelbracht et  al. in prep., Gordon et al. in
prep., Stansberry et al. in prep.). Color corrections  are also applied to MIPS photometry  for 
a power law spectrum with an index of 1.0.

The measurements are summarized in Tables 1 - 3. Notes to the tables give 
details regarding the aperture, extended source and color corrections.

\subsection{Radio Data}

A VLA  5GHz  radio   map  of  M87  was  obtained   at  a  resolution  of
$\sim$0.4$''$ on  1986 April  25 by \citet{Hines89}, who describe the observation
strategy  and data  reduction. 

\subsection{Image Convolution}

The  images  must be  convolved  to the  same  resolution  as MIPS  at
24$\mu$m   to  construct  the   full  spectral   energy  distributions
(SEDs). Therefore, we convolved the {\it  HST} and the 6 cm radio data
with  the MIPS  PSF. For  the 6  cm image,  we subtracted  the nuclear
emission by masking  the central ${1\farcs6}{\times}{1\farcs6}$ region
before convolution.  The IRAC images were convolved with kernels that
transform the observed IRAC PSFs to  the MIPS 24 $\mu$m PSF (Gordon et
al.,   in  prep.)   The  noise   images  were   calculated   by  error
propagation. The centroids of all images were aligned to the center of
the 6 cm data at RA(J2000)=12h30m49.42s, DEC(J2000)=12d23m27.97s.

\section{RESULTS}

\subsection{Infrared Image}

Figure~\ref{IMG_24res} shows the radio,  MIPS 24 $\mu$m, IRAC 8 $\mu$m
and optical image of M87 with nuclear subtraction at the resolution of
MIPS at 24  $\mu$m.  The stellar emission was  also subtracted for the
IRAC and  optical images. It is obvious  that the jet, Np  and Sf lobe
are all detected at 8 $\mu$m and 24 $\mu$m, with morphology similar to
the radio emission.  The infrared jet extends from the nucleus along a
straight line until it bends toward the southeast near the boundary of
the  Np  lobe.  The infrared  jet  peaks  at  the position  of  knot-A
(although not all knots are  resolved), which is the brightest knot in
the  radio  and  optical.   The  brightness  profile  across  the  jet
indicates  the width  of the  jet  is not  resolved at  24 $\mu$m  but
marginally resolved at 8 $\mu$m,  consistent with the width of the jet
of around  1$''$ at knot  A in the  optical image.  As  illustrated by
contours in  Figure~\ref{IMG_24res}, unlike  in the radio  and optical
image where  only one flux peak is  present in the Sf  lobe, there are
two flux peaks near the boundary  of the Sf lobe in the infrared.  The
radio  filaments  \citep[See][]{Hines89}  corresponding to  these  two
peaks are the filaments $\theta$ and $\eta$.

\begin{figure}
\epsscale{1.}
\plotone{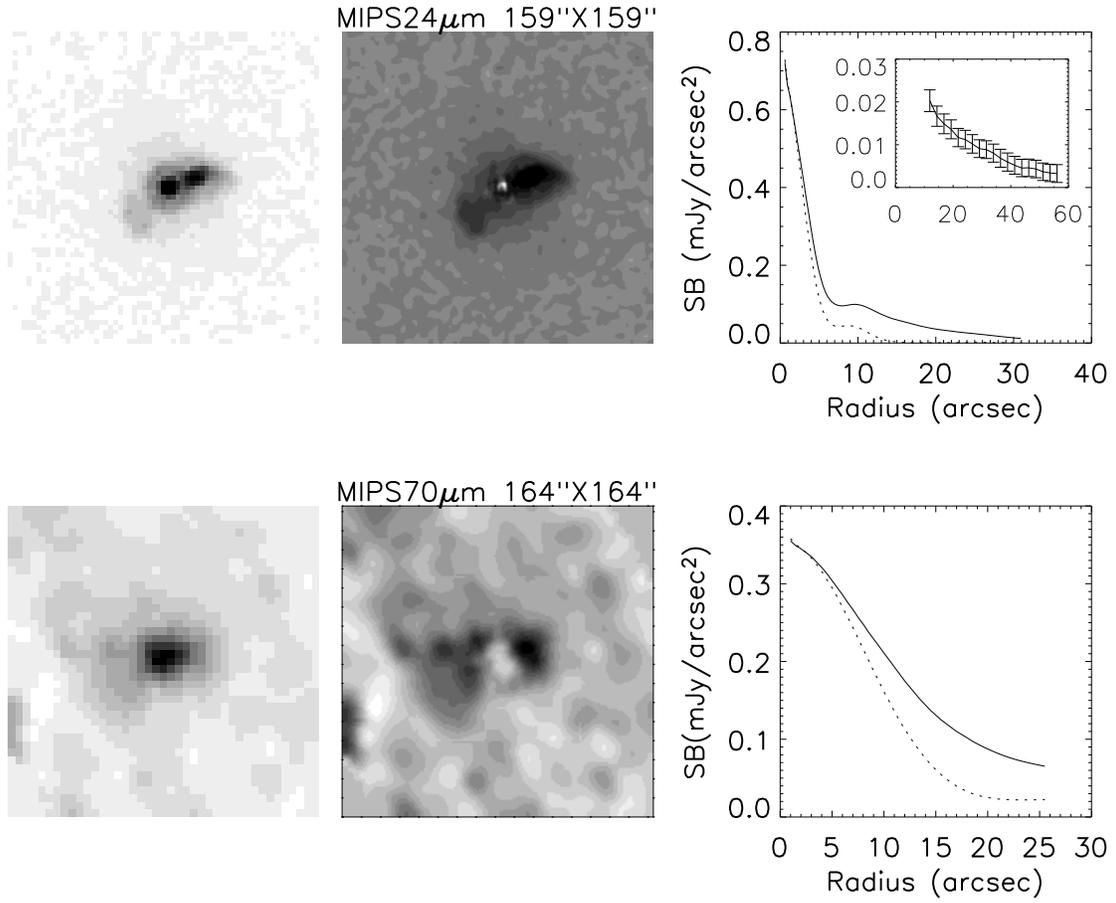} 
\caption{\label{MIPS} 
Original MIPS images of  M87 in the left column. MIPS images  of M87 after subtracting  
the nucleus in  the middle column. The radial profile of surface brightness is shown in 
the right column where the solid  line is the  profile of the  original image and  the dotted
line is the profile of the PSF at the corresponding wavelength.  The  subplot at 24$\mu$m 
shows  the surface brightness of the  emission after  masking the  central lobe region  defined by  the lowest
level contour in Figure 4 of \citet{Hines89}. }
\end{figure}

\begin{figure}
\epsscale{1.}
\plotone{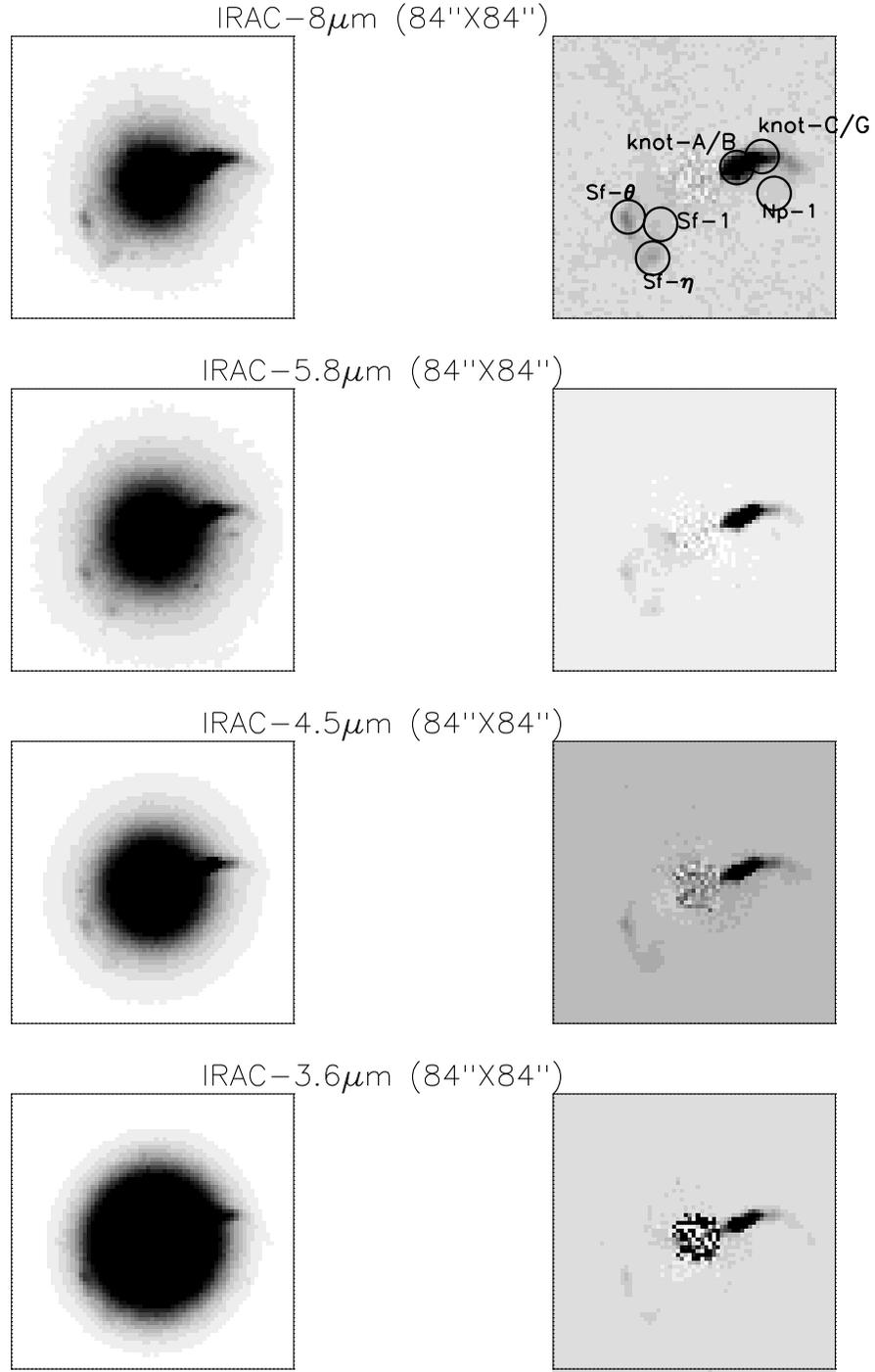} 
\caption{\label{IRAC}  
Original {\it IRAC} images  in the  left column and galaxy-light
subtracted images in the right column.}
\end{figure}

\begin{figure}
\epsscale{1.}
\plotone{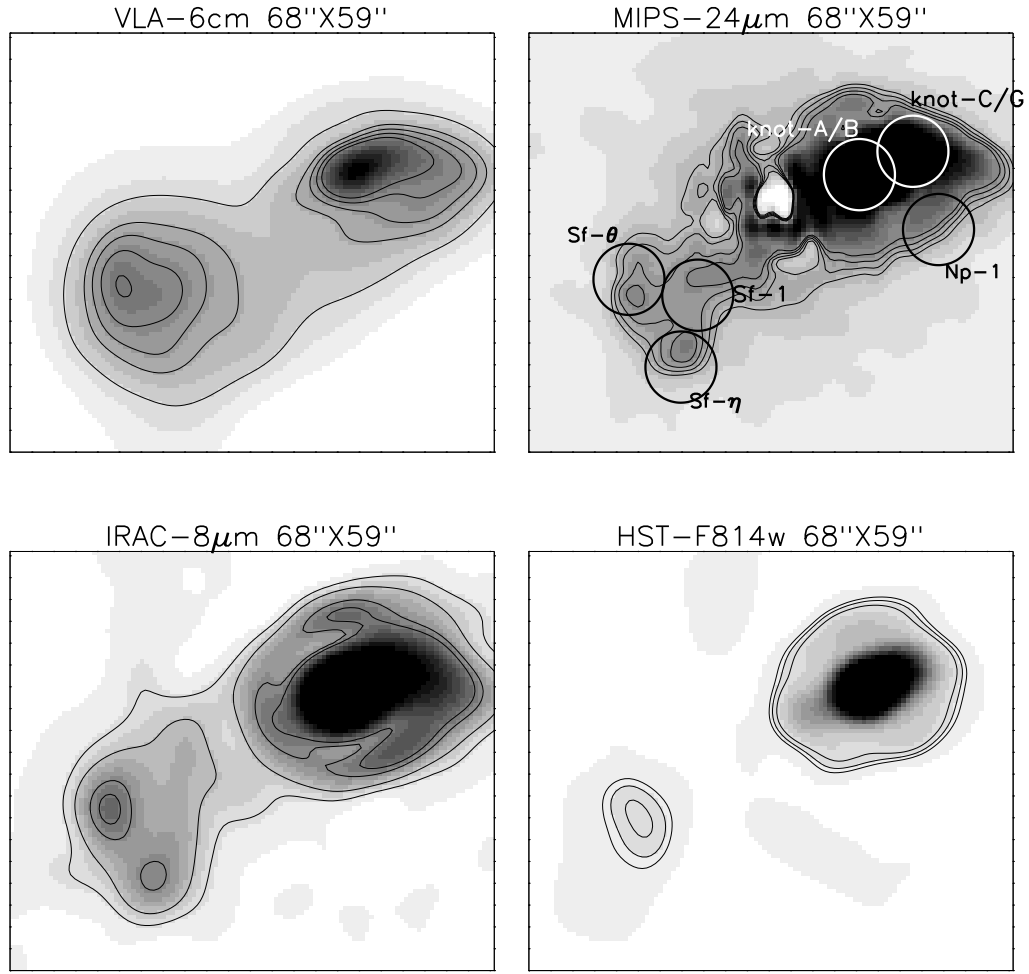} 
\caption{\label{IMG_24res} 
The radio, MIPS 24$\mu$m, IRAC  8$\mu$m and {\it HST} optical image of
M87  after subtracting  the  nucleus  at the  resolution  of the  MIPS
24$\mu$m image.  The galaxy light  is also subtracted for the IRAC and
{\it HST}  images. The contours  (thin solid lines) are  superposed to
illustrate two local maxima in  the Sf lobe at the infrared wavelength
while only one maximum is present in the radio and optical.}
\end{figure}

\begin{figure}
\epsscale{1.}
\plotone{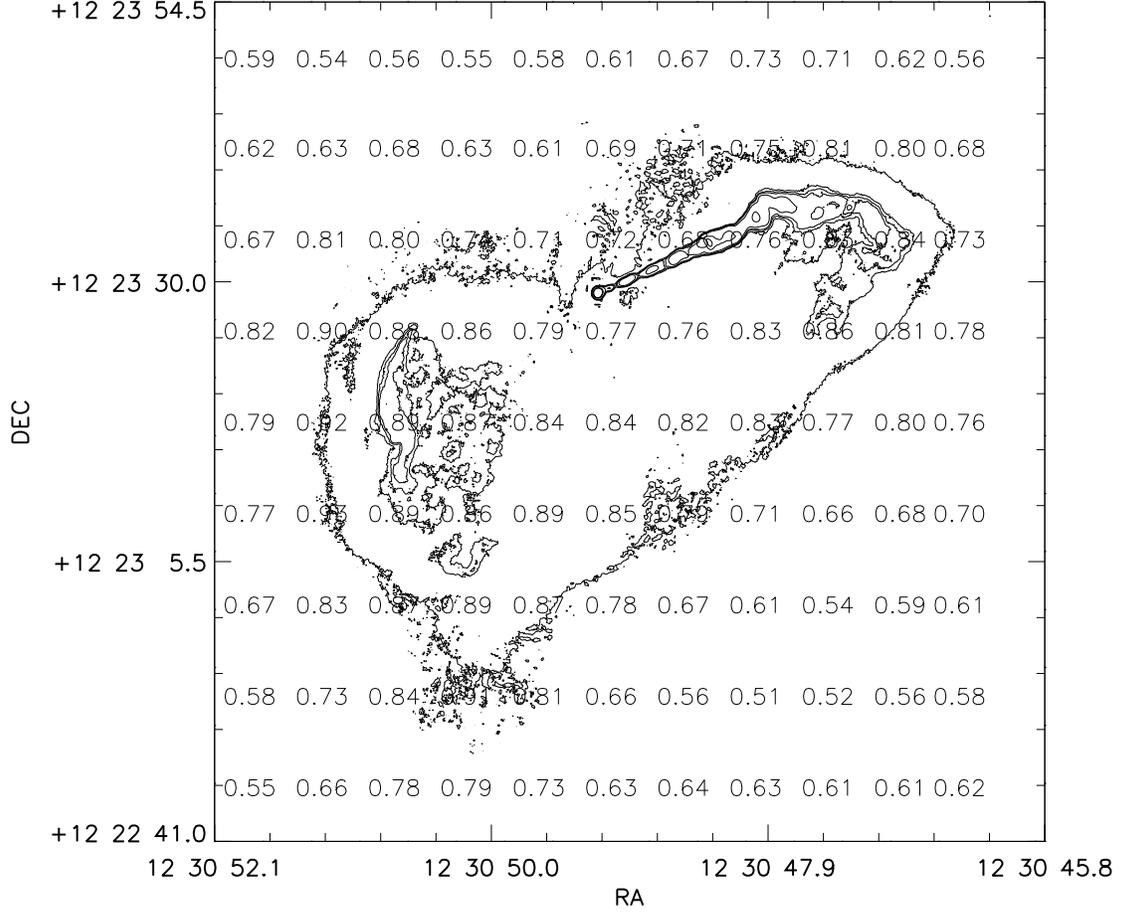}
\caption{\label{spec_index} 
The  map   of  the  spectral   index  $\alpha_{5GHz}^{24{\mu}m}$.  The
measurements of  flux density at 5GHz  and 24 $\mu$m were  made on the
image  with  nuclear subtraction  at  the  resolution  of MIPS  at  24
$\mu$m. The solid contour is  the intensity distribution of the original 6
cm image. Contour  intervals are [0.11, 0.43, 0.53,  0.64, 1.06, 5.32,
10.64, 173.86] mJy.}
\end{figure}

Besides  the infrared jet  and lobes,  an infrared  halo is 
clearly visible at  24 $\mu$m.   The  subplot in
Figure~\ref{MIPS} shows  the surface brightness of the  emission at 24
$\mu$m after  masking the  central lobe region  defined by  the lowest
level contour in Figure 4 of \citet{Hines89}.  Unlike the lobe, it has
a  spherical  structure  as  shown  in  Figure~\ref{MIPS},  with  flux
decreasing  with  increasing  distance  from the  nucleus.   Moreover,
although the  infrared halo is  on the same  scale as the radio inner region
(${1\farcm5}{\times}{2\farcm0}$),  the  morphologies  are  quite
different since the radio inner region is elongated and aligned with the
radio lobe  \citep[See][]{Owen00}.  The discrepancy  in the morphology
implies that the emission of the infrared halo arises from a different
mechanism. Figure~\ref{spec_index} shows the map of the spectral index
$\alpha_{5GHz}^{24{\mu}m}$ where  the photometry at each  grid point at 5 GHz
and 24  $\mu$m was carried out on  the image at the resolution of MIPS at 
24 $\mu$m and with the nucleus subtracted.    The    halo    has
$\alpha_{5GHz}^{24{\mu}m}$  flatter than  the nonthermal  emission region,
again suggesting a different emission mechanism.

As shown in  Figure~\ref{MIPS}, at 70 $\mu$m, the jet  and Np lobe are
detected  but are  blended;  the Sf  lobe  is also  detected.  At  160
$\mu$m,  the  jet and  lobes  are all  blended  at  the {\it  Spitzer}
resolution. The infrared  halo is not detected at  either 70 $\mu$m or
160 $\mu$m.

\subsection{ Infrared Emission from the Host Galaxy}

Table~\ref{photo_tot}  shows  photometry of  M87  within  a radius  of
30$^{\prime\prime}$,  including emission  from the  nucleus,  jet, and
most of the Np and Sf lobes. The photometry for the Two Micron All Sky
Survey (2MASS) was carried out  on the original images without nuclear
subtraction.  The ISOCAM  photometry is  from  \citet{Xilouris04}; the
ISOCAM measurements at shorter wavelengths have been supplanted by the
IRAC  data, which  has a  substantially  improved ratio  of signal  to
noise. We also include IRAS  Faint Source Catalog (FSC) data. Since it
was  extracted using a  point source  filter, the  effective apertures
should be  close to  the 60$^{\prime\prime}$ used  for the  other data
sources.

To interpret these data, we fitted a stellar SED in the near infrared,
taking  galaxy colors from  \citet{Johnson66} (because  this reference
consistently  integrates JKL  -  we applied  color  corrections as  in
\citet{Carpenter01} and Rieke et al., in prep.). The SED was continued
to the longer IRAC bands  using the measurements of individual K stars
in   \citet{Reach05}.   In  normalizing   the  stellar   spectrum,  we
emphasized the IRAC 3.6 and 4.5$\mu$m points because systematic errors
within the IRAC  measurements should be smaller than  between them and
data from other sources. Specifically, the photometry in all four IRAC
bands  was  conducted on  the  SSC  post-BCD  images, using  the  same
apertures and sky reference areas.
 
We  fitted two  power laws  to  the nonthermal  spectrum, because  the
nucleus  is   self-absorbed  at     frequencies  below   $\sim$2  GHz
\citep[e.g.][]{Charlesworth82} and  hence has a  different spectrum from
the  other  components.   The  nuclear  component  is  fitted  to  the
10.8$\mu$m measurement  of \citet{Perlman01b}, our  measurements at 24
and  70$\mu$m,  and  those  reported  by  \citet{Haas04}  at  450  and
850$\mu$m.  It  has $\alpha  = 1.15$  and a flux  density of  45mJy at
24$\mu$m.   The remaining nonthermal  emission remains  optically thin
into the radio,  down to 1.4GHz.  It has been fitted  with a power law
with $\alpha = 0.92$ and a flux density of 50mJy at 24$\mu$m.  Because
the break frequency for the  majority of the nonthermal emission is in
the optical or  UV (Section 3.3.3), we continue  the power laws toward
shorter wavelengths.

\begin{figure}
\epsscale{1.}
\plotone{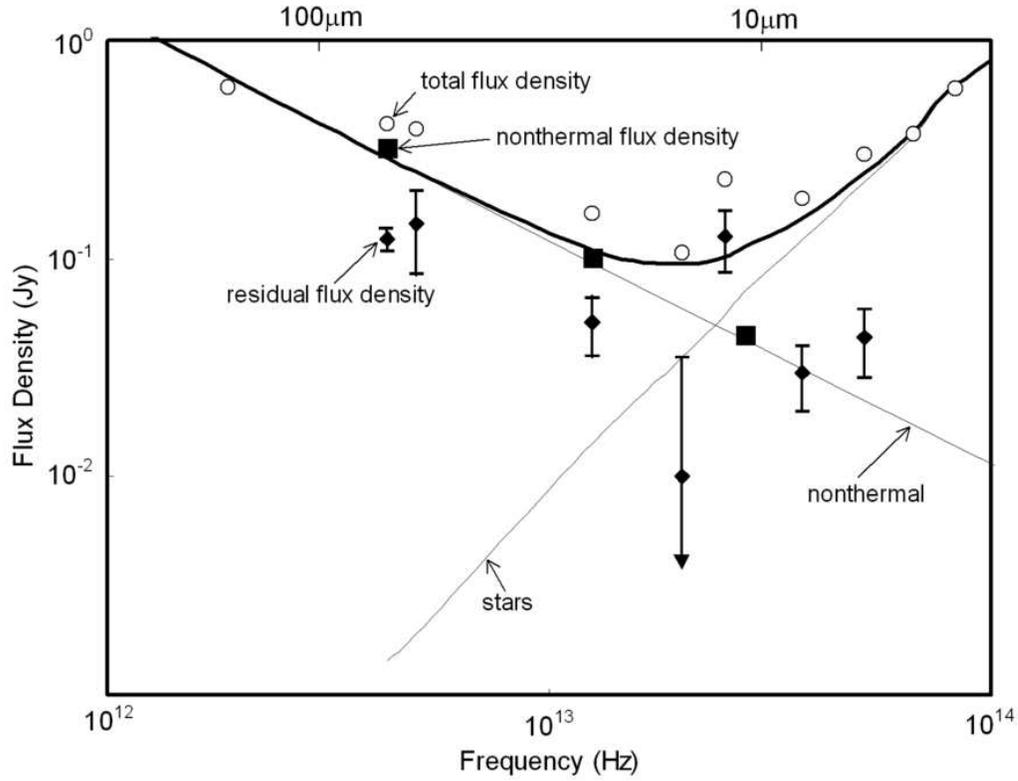}
\caption{\label{sed_full}
Comparison of large aperture photometry with models
of the stellar and nonthermal SEDs. The light lines show the models
of the power law nonthermal emission and the stellar output, and the
heavy line is the sum of the two components. The open circles
are the 60 arcsec diameter aperture photometry. The filled boxes
show the flux densities for the nucleus, jet, and lobes
from this work at 70 and 24$\mu$m, and from \citet{Perlman01b} (with
a correction for low surface brightness components) at
10.8$\mu$m. The diamonds show the residual fluxes after
subtracting the model from the totals measured. At 70$\mu$m, we label
the points for total, nonthermal, and residual flux density.}
\end{figure}

Table~\ref{photo_tot} shows  the aperture photometry,  the stellar and
nonthermal spectral  components, and the residual  flux densities when
the  modeled   SEDs  are  subtracted  from   the  measurements.   This
information is also illustrated  in Figure~\ref{sed_full}.  There is a
well detected excess above the model from 6 through 70$\mu$m, which is
particularly  strongly  indicated at  24  and  70$\mu$m. However,  the
measured values for the  nonthermal emission (nucleus, jet, and lobes)
at  these two  wavelengths  fall very  close  to the  model, which  is
dominated by the nonthermal  emission.  Therefore, it appears that the
excess  is not associated  with the  resolved nonthermal  sources. The
possibility  that  this  emission  is  associated  with  the  extended
nonthermal radio  emission can  be tested from  Figure 5,  which shows
that the spectrum between 5GHz and 24$\mu$m becomes {\it flatter} with
increasing distance  from the nucleus  and jet.  This behavior  is not
expected for extended nonthermal  emission, which should either retain
the  same slope or  have a  steeper one  due to  energy losses  in the
synchrotron electron spectrum.

\citet{Sparks93}  discuss  filaments  visible  in  H$\alpha$.   It  is
plausible  that  dust  in  these  filaments contributes  some  of  the
infrared emission  not associated directly with  the nonthermal source
components.    However,   the  morphology   of   the  24$\mu$m   image
(Figure~\ref{MIPS}) shows  it to be  more symmetric and  more extended
than  the H$\alpha$ filaments,  which are  concentrated in  the region
occupied  by the  jet  and  lobes.  Therefore,  most  of the  infrared
emission must arise from a different component of the galaxy. The 
agreement with  the  pure elliptical  isophotes  seen in  the
optical  surface brightness  \citep{Liu05} indicates  the  dust should
be smoothly distributed in the host galaxy.

The overall level of excess emission  at 70$\mu$m is at a normal level
for  giant elliptical  galaxies  \citep[e.g.][]{Temi04,Leeuw04}, where
the dust is  heated by the stellar light  or collisions with electrons
in the  hot gas.   The overall SED  of M87  from the near  infrared to
24$\mu$m  is  very  similar   to  those  of  the  low-activity  (i.e.,
non-infrared-luminous) brightest cluster  galaxies (BCGs) discussed by
\citet{Egami06}.  The excesses  detected weakly  in the  IRAC  5.8 and
8$\mu$m bands and in the  IRAS FSC at 12$\mu$m suggest the possibility
of  weak, extended aromatic  emission, which  is seen  in a  number of
X-ray-emitting elliptical  galaxies by \citet{Kaneda05}  (but see also
\citet{Bregman06}).   However, aromatic  emission is  not seen  in the
nuclear  spectrum  \citep{Bressan06},  which  instead  shows  silicate
emission   either   associated   with   the  AGN   or   with   evolved
stars. Therefore, if  these features are present they  must arise from
dust  distributed within  the galaxy  and perhaps  heated as  in other
X-ray-emitting giant ellipticals.  In any case, the characteristics of
the mid- and far-infrared excess above the compact nonthermal emission
from M87  appear to be consistent  with expectations for  a normal BCG
and need have little to do with the nonthermal activity in the nucleus
and jet.

\subsection{Behavior of the Nonthermal Sources}

\subsubsection{SEDs of the Nonthermal Sources}

\begin{figure}
\epsscale{1.}
\plotone{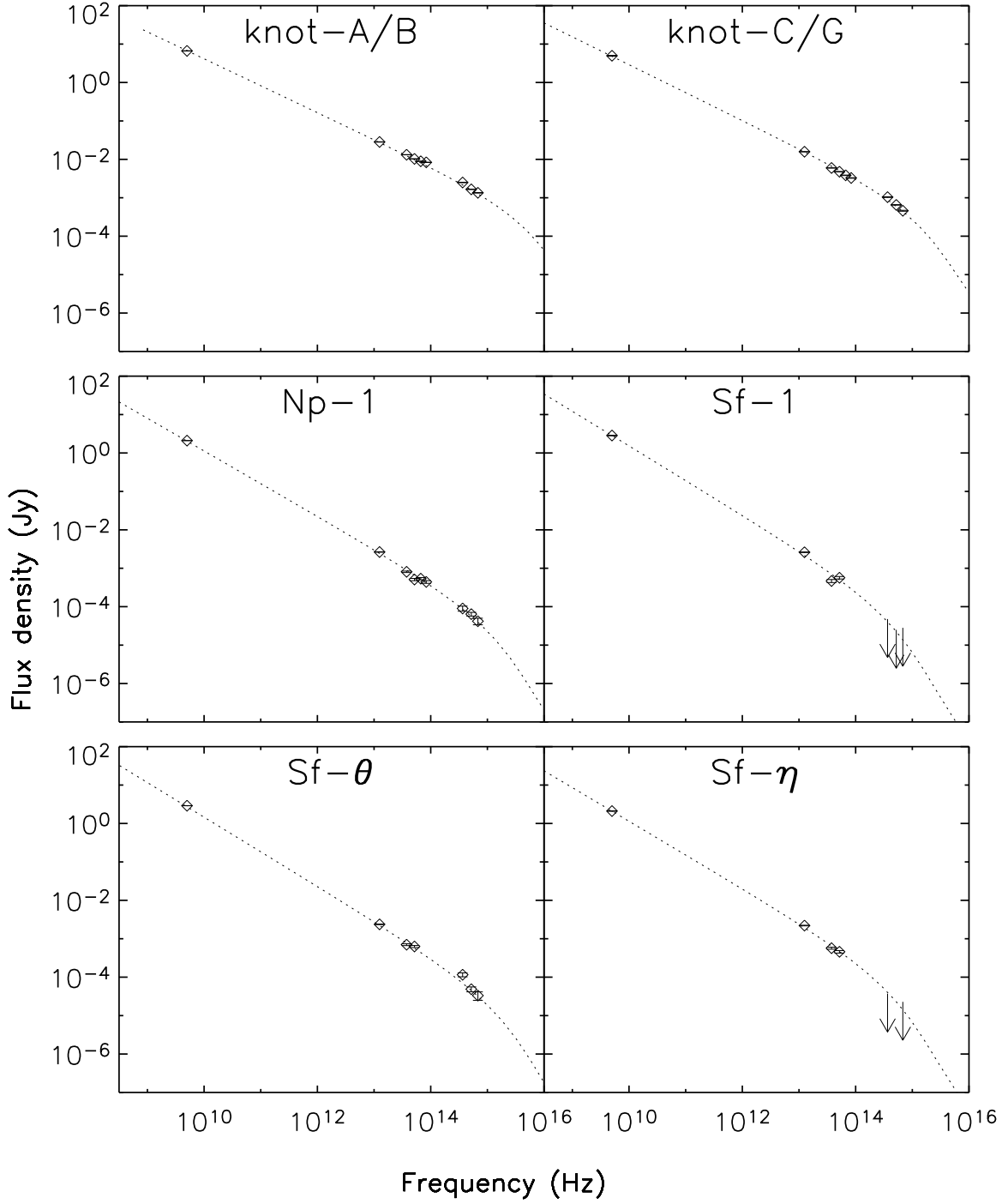} 
\caption{\label{fit_SED} 
SEDs  of  individual   regions  in  the  MIPS  24   $\mu$m  image  (see
Figure~\ref{MIPS}).  The photometry of the nonthermal features is shown as diamonds,
with downward pointing arrows for the upper limits.
The lines are the fits of the KP synchrotron emission models. }
\end{figure}

We  now consider the  extranuclear nonthermal  emission.  As  shown in
Figure~\ref{fit_SED}, we constructed the SED from the radio to optical
wavelengths for two regions in the jet (knot-A/B and knot-C/G
\footnote{we use  composite designations  to emphasize that  the knots
are blended at  our resolution.}, one region (labeled  as Np-1) in the
Np lobe, and three regions  (Sf$-$1, Sf$-\theta$ and Sf$-\eta$) in the
Sf lobe.  Each region is circular with a radius of 5$''$. The position
of each region and the corresponding 24 $\mu$m flux density are listed
in Table~\ref{photo_24um}.

Figure~\ref{fit_SED}  shows the  SEDs of  the different  regions.  The
 stellar  emission is modeled with a blackbody spectrum
fitted to the total IRAC  3.6 $\mu$m and total optical emission within
each region.    Table~\ref{photo_24um} lists the
radio-infrared  spectral index  $\alpha_{R}^{24{\mu}m}$ between  5 GHz
and  24$\mu$m  after  subtracting  the stellar  contribution  and  the
radio-optical  index  $\alpha_{R}^{opt}$ between  5GHz  and the  three
$HST$ bands. In the jet and lobe, it seems that the regions with lower
infrared  surface brightness  have a  steeper $\alpha_{R}^{24{\mu}m}$.
$\alpha_{R}^{24{\mu}m}$  is also  flatter than  $\alpha_{R}^{opt}$, as
expected  for synchrotron  radiation with  a higher  rate  of electron
energy   loss  at   the  higher   energies.   The   larger  steepening
${\Delta}{\alpha}=\alpha_{R}^{opt}-\alpha_{R}^{24{\mu}m}$  in the lobe
($\sim$0.03-0.1)  than  the  jet  ($\sim$0.01-0.03) reveals  that  the
high-energy cutoff  of electron energy in  the lobe occurs  at a lower
frequency, as demonstrated by the SED fits below.

\subsubsection{Minimum Pressure Analysis}

Assuming the equipartition condition that the particle energy is equal
to  the  magnetic  energy,  the  minimum magnetic  field,  energy,  and
pressure of the emitting plasma  can be derived from the observed SED,
the    emitting    volume    and    the    filling    factor    $\phi$
\citep{Pacholczyk70}.  The standard  minimum pressure analysis assumes
a power law distribution of  electron energy; the dependence of
the physical  parameters on the  high frequency cut-off  is negligible
 if  the spectral index $\alpha$  $>$ 0.5, as for M87.  This  implies that the
synchrotron loss at high frequencies  does not change
the results for M87.   

During  the calculation, the ratio of proton  energy to electron energy is assumed
to be  unity.  The SED  is assumed  to be a  power law with  a spectral
index equal to $\alpha_{R}^{24{\mu}m}$ between 10$^8$ Hz and the break
frequency. A low  frequency cut-off of 10$^8$ Hz  was adopted, as the
SED  of jet  and  lobe  starts to  deviate  from a power  law  at 1  GHz
\citep{Felten68, Meisenheimer96}. One  order of magnitude variation in
the  low frequency  cutoff  results in  a  factor of  $\sim$1.2 change in  the
magnetic field and  a factor of $\sim$1.4 change in  the energy and pressure.
The volume  is taken to be spherical with  a radius  of 5 arcsec  and the
filling factor  is assumed to be 0.1. A factor of  10 variation in
the filling  factor corresponds  to a factor  of $\sim$2 in  the magnetic
field, energy, and pressure.  The filling  factor is hard to determine
as it depends  on the true three dimensional  structure of the synchrotron
plasma.   The value  of  0.1 was  adopted  to balance  among
various alternatives, such  as a whole filled volume or a thin
boundary   layer  of   the  volume   as  proposed   by   some  studies
\citep{Owen89}. 

The last three columns of Table~\ref{spec_fit} list the minimum magnetic field, energy,
and  pressure  computed with  formulae  given by  \citet{Burns79}. Our
results agree well with those of \citet{Eilek03}.  Independent support 
for the derived magnetic field strengths comes from
the inverse Compton fluxes measured in Tev gamma rays. \citet{Stawarz05} 
analyze the HESS and HEGRA data to derive a lower limit of 300$\mu$G. 
However, the Tev gamma ray limit cannot be applied individually to the
various regions we have isolated in M87.

\subsubsection{Model of the Synchrotron Emission}

We employ  a quantative model to extract the underlying physical parameters
regulating  the   SED.  The   Kardashev-Pacholczyk    (KP)   model
\citep{Kardashev62,  Pacholczyk70} has been shown to fit the jet behavior
well \citep{Perlman01a}. It describes the  synchrotron emission
of  an ensemble  of  electrons  ejected $t$  seconds  ago. Assuming  the
initial  energy distribution  of the electrons  is a  power  law $N(E_{0},
\omega, 0)=N_{0}E_{0}^{-\gamma}$  where $\omega$  is the pitch  angle, the
new energy  distribution of electrons  at time $t$ due  to synchrotron
loss is:
\begin{equation} 
 N(E, \omega, t) = N_{0}E^{-\gamma}(1 - c_{2}B^{2}sin^{2}(\omega)Et)^{\gamma-2}
\end{equation}
where $c_{2}=2.37\times10^{-3}$ (in cgs units) as defined in \citet{Pacholczyk70}. Integrating over the pitch angle,
the intensity of the synchrotron emission from these electrons at frequency $\nu$ is:
\begin{equation}
  I_{\nu} = 2{\pi}sc_{3}c_{1}^{(\gamma-1)/2}N_{0}B^{(\gamma+1)/2}\tilde{\nu}^{(1-\gamma)/2}_{T}\tilde{B}(\tilde{x}_{T}, \gamma)
\end{equation}
where $s$  is the  extension of  the source along  the line  of sight,
 $c_{1}$=6.27$\times10^{18}$ (in cgs units), $c_{3}$=1.87$\times10^{-23}$ (in cgs units),  $B$  is the magnetic  field,
$\tilde{\nu}_{T}=c_{1}/(c_{2}^2B^{3}t^{2})$.  and   $\tilde{x}_{T}   =
\nu/\tilde{\nu}_{T}$.   Therefore,  for a  given  magnetic field,  the
shape  of  the synchrotron  radiation spectrum  is  described  by  three  physical
parameters:  the break frequency (  $\tilde{\nu}_{T}$ ), the initial power-law
index   of  the   electron  energy   distribution  ($\gamma$)   and  the
normalization (flux at a given frequency).

The dotted line in Figure~\ref{fit_SED}  shows the SED model fitted to
the photometry excluding upper limits.   The results from the SED fits
for regions in the jet and lobe are given in Table~\ref{spec_fit}.  As
a check  of our calculations,  the break frequency in  knot-A/B agrees
well with previous  work; for example, it is  intermediate between the
values for the individual knots  A and B obtained by \citet{Waters05}.
Our values for the knots are $\ge  1-5 \times 10^{15} $ Hz.  In the Np
and Sf lobes, the turnover of  the spectrum occurs in the optical with
a break frequency at least a  factor of five lower than for knots A/B,
except  for  the Sf$-\theta$  region,  which  has  a comparable  break
frequency to the jet.  \citet{Stiavelli97} obtained a cutoff frequency
of 4.3$\times10^{14}$ Hz for the feature $\theta$, almost one order of
magnitude lower than our value.  The discrepancy arises mainly because
of the difference  in the optical photometry of  the feature $\theta$.
We  believe that the  high resolution  and high  S/N {\it  HST} images
should  minimize   the  uncertainties  associated   with  the  stellar
subtraction.   Nonetheless,   accurate  determination  of   the  break
frequencies requires photometry in the UV, which is not used in either
study. However, our conclusions in this paper depend only on the break
frequencies lying at optical or  shorter wavelengths, not on the exact
frequencies involved.

The    lowerlimit   of   the    current   maximum    electron   energy
\citep{Uchiyama06} derived from the break frequency
\begin{equation}
 E_{max}                                                              =
 0.28(\frac{1+z}{\delta})^{0.5}(B/10^{-4})^{-0.5}(v_{t}/10^{14}
 )^{0.5} ~{\rm Tev}
\end{equation}
is  another physical  parameter constraining  the  acceleration model,
where  $z$ is the  redshift and  $\delta$ is  the Doppler  factor. For
$z$=0  and $\delta$=1,  as  shown in  Table~\ref{spec_fit}, a  maximum
electron energy with Lorentz  factor greater than 10$^{6}$ prevails in
both the jet and lobe regions.

\subsubsection{{\it In Situ} Reacceleration in the Lobe}

The {\it Spitzer} data  provide improved constrains on the synchrotron
spectrum  in the Np  and Sf  lobes, showing  that the  break frequency
occurs in the optical or in the UV for Sf$-\theta$.  The corresponding
synchrotron  age  \citep[$t=(c_{1}/(  c_{2}^{2}B^{3}{\nu}_{t}))^{0.5}$
secs;][]{Pacholczyk70}  for the  two  jet regions  and Sf$-\theta$  is
around  500  yrs   while  electrons  in  the  Np   and  Sf  lobes  are
$\sim$1$\times10^{3}$  yrs old.  The  largest distance  that electrons
can travel in 1000 yrs must be smaller than $ct$ = 300pc (4$''$) where
$c$ is the speed of  light. Therefore, {\it in situ} reacceleration of
electrons  must  occur  in  the  jet \citep[as  found  previously  by,
e.g.,][]{Perlman01a,  Eilek03, Waters05}, and  in both  the Np  and Sf
lobes.   \citet{Hines89} showed  how this  process could  explain some
additional features in the lobes, such as the bright radio filaments.

As shown in Table~\ref{spec_fit},  the individual regions in the lobes
have  a  minimum electron  energy  around  10$^{54}$  erg and  thus  a
conservative  estimate of  the electron  energy in  an entire  lobe is
around 10$^{55}$ erg, considering  its probable total volume.  Given a
synchrotron age of 10$^{3}$ yrs, the minimum input energy rate into an
entire lobe is around 3$\times$10$^{44}$ erg s$^{-1}$.  Assume for the
moment that this energy is  supplied by supernovae.  If each supernova
releases an energy of 10$^{52}$ erg, the explosion rate accounting for
the  observed input  energy  rate  is 1  yr$^{-1}$,  much higher  than
observed in  normal galaxies.  The gravitational  potential as implied
by the X-ray luminosity of thermal  gas is much lower compared to this
reacceleration  rate.  The most  plausible energy  source is  the jet,
especially for the Np lobe where  the jet is visible.  The minimum jet
power  is around 3$\times$10$^{44}$  erg s$^{-1}$  \citep{Owen00}.  If
the  electrons  in the  lobe  are reaccelerated  {\it  in  situ} by  a
mechanism driven by  the jet, most of the jet  power must be converted
into the kinematics of electrons in the lobe.

Radial  outflow  connecting the  lobe  and  outer  halo is  frequently
interpreted   as  a   buoyant  bubble   originally  inflated   by  jet
\citep{Owen00};  the radio  lobe may  be inflating  the  bubble before
detaching from the nucleus \citep{Gull73, Bicknell96, Churazov01}. The
rising  bubble captures  the hot  X-ray gas  from the  ambient medium,
creating  the X-ray  arc as  seen in  the X-ray  image \citep{Young02,
Forman06}.  In general, the total  energy of jet power is converted to
the internal  energy of the  lobe or to  the work to inflate  the lobe
\citep[e.g.][]{Bicknell97}. The relative fraction may be determined by
the expansion process.  The result of this study, that most of the jet
power is locked into the kinetic  energy of the electrons in the lobe,
implies that only a small fraction of the jet power is used to inflate
a cavity (radio  lobe) in the thermal gas  atmosphere. This conclusion
appears to  be consistent with  the X-ray observation that  the energy
for   inflation   is  much   lower   than   the   minimum  jet   power
\citep[See][]{Young02}.   The   kinetic  energy  of   the  high-energy
electrons  should  be carried  away  by  radiation  due to  the  short
synchrotron lifetime, while the radio plasma could live long enough to
inflate  the lobe  if synchrotron  radiation is  the only  energy loss
mechanism.      As     shown     in     Table~\ref{photo_24um}     and
Table~\ref{spec_fit}, the total synchrotron  luminosity of the lobe is
much smaller than  the energy rate for electron  reacceleration. It is
possible  that  most  of  the low-energy-electron  kinetic  energy  is
carried away by bubbles to the outer halo to heat the cooling flow.

\section{Conclusions}

We present {\it Spitzer} observations of the giant elliptical galaxy 
M87 in the Virgo cluster.  The basic conclusions are:

(1) The far infrared excess emission above the non-thermal power law
is likely to be warm dust distributed throughout the galaxy. Its
luminosity is similar to that observed in the far infrared for
other, normal giant elliptical galaxies.

(2) The  infrared emission  in the  jet and  two lobes  is synchrotron
emission with break frequencies in the UV (jet) or optical (lobes).  
The high break frequency in the lobes indicates that {\it in situ} reacceleration 
of electrons occurs there as well as in the jet. The minimum  input
energy rate to reaccelerate the  electrons in the lobe is comparable to
the jet power, implying  that most of the jet power  is converted to electron
kinematics in the lobe, not used to inflate the lobe.

%\section{Acknowledgments}
\acknowledgements  
We wish  to  thank the  the  anonymous referee  for
detailed comments and Frazer Owen for providing the electronic version
of  the radio  image  from  \citet{Hines89}.  This  work  is based  on
observations made  with {\it  Spitzer}, which is  operated by  the Jet
Propulsion Laboratory, California  Institute of Technology, under NASA
contract 1407.   This work also made  use of images  obtained from the
data  archive  at the  Space  Telescope  Science  Institute. STScI  is
operated by the Association of Universities for Research in Astronomy,
Inc., under  NASA contract  NAS5-26555. Radio data  are from  the NRAO
Data Archive System, supported  by the National Science Foundation. We
also  used  the  NASA/IPAC  Extragalactic  Database  (NED),  which  is
operated  by the  Jet Propulsion  Laboratory, California  Institute of
Technology,  under contract  with the  National Aeronautics  and Space
Administration.   The  work was  supported  by  contract 1255094  from
JPL/Caltech to the University of Arizona.

\clearpage

\begin{deluxetable}{lllllrlll}
\tablecolumns{8}
\tabletypesize{\scriptsize}
\tablecaption{ \label{photo_tot} Photometry within the central 60$''$ diameter region}
\tablewidth{0pc}
\tablehead{\colhead{Wavelength}  &  \colhead{Total Flux Density}  &  \colhead{Stellar Component} &  
\colhead{Nonthermal Component} & \colhead{Residual} \\
\colhead{($\mu$m)}  &  \colhead{(mJy)}  &  \colhead{(mJy)}  &  \colhead{(mJy)}  &  \colhead{(mJy)} \\
}
\startdata
1.25       &  1241$\pm$124$^1$   & 1122 & 5   & --         \\
2.16       &  1202$\pm$120$^1$   & 1110 & 9   & --         \\
3.6        &  605$^2$            & 616  & 15  & --         \\
4.5        &  375$^2$            & 365  & 19  & --         \\
5.8        &  302$^2$            & 225  & 24  & 53$\pm$15$^6$ \\
8          &  189$^2$            & 121  & 33  & 35$\pm$10$^6$ \\
12         &  231$\pm$37$^3$     & 55   & 49  & 127$\pm$40 \\
15         &  106$\pm$21$^4$     & 35   & 62  &  9$\pm$25  \\
24         &  171.9$\pm$13.4$^5$ & 14   & 99  & 59$\pm$15  \\
60         &  394$\pm$51$^3$     & 2    & 251 & 141$\pm$60 \\
70         &  455.2$\pm$9.2$^5$  & 1    & 293 & 161$\pm$15 \\
160        &  581.5$\pm$10.0$^5$ & 0    & 682 &     -      \\

\enddata
\tablecomments{$^1$2MASS; $^2$IRAC, from PID 3228, PI W. Forman. We used extended source corrections of
0.934 at 3.6$\mu$m, 0.97 at 4.5$\mu$m, 0.86 at 5.8$\mu$m and 0.80 at 8$\mu$m; $^3$IRAS FSC; 
$^4$ \citet{Xilouris04}; $^5$MIPS, from PID 82, PI G. Rieke, aperture corrections of 1.09, 1.31 and 1.93 respectively
at 24, 70 and 160 $\mu$m,  color corrections of 1.04, 1.1 and 1.04 respectively at 24, 70 and 160 $\mu$m; 
$^6$Errors are taken to be 5\% of the total flux density in the band.}
\end{deluxetable}

\begin{deluxetable}{lllllrlll}
\tablecolumns{8}
\tabletypesize{\scriptsize}
\tablecaption{ \label{photo_70um} Photometry at 70 $\mu$m of the nucleus, jet and lobes}
\tablewidth{0pc}
\tablehead{ \colhead{Region}  &  \colhead{RA}   & \colhead{DEC}  
            & \colhead{ $f_{70{\mu}m}$(Jy) }    }
\startdata
Nucleus              &     12 30 49.42 & 12 23 27.97  &   228$^{1}$    \\
Jet \& Np lobe       &     12 30 48.21 & 12 23 33.9   &   70$^{2}$     \\
Sf lobe              &     12 30 50.49 & 12 23 14.2   &   57$^{2}$     \\
\enddata
\tablecomments{ $^{1}$From PSF fitting;
$^2$Aperture correction factor of 1.3 for a 30$''$-diameter aperture; Color correction factor of 1.1 for all three fluxes.}
\end{deluxetable}

\begin{deluxetable}{lllllrllllll}
\tablecolumns{8}
\tabletypesize{\scriptsize}
\tablewidth{0pc}
\tablecaption{ \label{photo_24um} Spectral properties of the nucleus, jet and lobes at the resolution of MIPS at 24$\mu$m}
\tablehead{       \colhead{Region} & \colhead{RA} &  \colhead{DEC} &  \colhead{ $f_{24{\mu}m}$ } &  \colhead{ $f_{8{\mu}m}$ } 
&  \colhead{ $f_{5.8{\mu}m}$ } &  \colhead{ $f_{4.5{\mu}m}$ } &  \colhead{ $f_{3.6{\mu}m}$ } & \colhead{$\alpha_{R}^{24{\mu}m}$} 
                                & \colhead{$\alpha_{R}^{opt}$}  &  \colhead{$L_{tot}^{syn}$} \\
                  \colhead{} & \colhead{}&  \colhead{} &  \colhead{(mJy)}                  & \colhead{(mJy)} 
                                & \colhead{(mJy)}                 & \colhead{(mJy)}                   & \colhead{(mJy)}
                                & \colhead{}          & \colhead{}     & \colhead{(erg)}  }
\startdata
      Nucleus      &     -       &    -        &      50.7$^{1}$&        - &       -  &        - &        - &         -  &            -  &   -       \\
    knot$-$A/B     & 12 30 48.64 & 12 23 32.2  &      32.1$^{2}$&     13.19&     10.29&      8.92&      8.36&        0.70&           0.71&  1.81E+42 \\
    Knot$-$C/G     & 12 30 48.12 & 12 23 35.4  &      17.6$^{3}$&      5.97&      4.77&      3.84&      3.27&        0.74&           0.77&  6.88E+41 \\
    Np$-$1         & 12 30 47.88 & 12 23 24.5  &      3.22$^{4}$&      0.81&      0.51&      0.54&      0.44&        0.85&           0.91&  1.03E+41 \\
    Sf$-$1         & 12 30 50.19 & 12 23 15.3  &      3.45$^{4}$&      0.47&      0.57&        - &       -  &        0.89&    $>$    0.99&  9.29E+40 \\
    Sf$-\theta$    & 12 30 50.85 & 12 23 17.5  &      2.93$^{4}$&      0.70&      0.63&        - &       -  &        0.91&           0.94&  1.01E+41 \\
    Sf$-\eta$      & 12 30 50.35 & 12 23  5.2  &      2.70$^{4}$&      0.57&      0.46&        - &       -  &        0.88&     $>$   0.98&  7.85E+40 \\
\enddata
\tablecomments{ $^1$ Nuclear flux density obtained by fitting PSF to the surface brightness profile; 
$^2$An aperture correction factor of 1.76; $^3$An aperture correction factor of 1.40; 
$^4$No aperture correction; Color correction factor of 1.04 at 24$\mu$m.}
\end{deluxetable}

\begin{deluxetable}{llrrrlllll}
\tablecolumns{8}
\tabletypesize{\scriptsize}
\tablewidth{0pc}
\tablecaption{ \label{spec_fit} The physical parameters from the spectrum fit}
\tablehead{        \colhead{Region} &    \colhead{$\alpha_{INPUT}$}   &   \colhead{$\nu_{t}$(Hz)} &   
         \colhead{$t_{syn}$(yr)} &    \colhead{$Ee_{max}$(0.5Mev)} &   \colhead{$B_{minP}$(G)}    & 
         \colhead{$E_{minP}$(erg)}    &    \colhead{$P_{minP}$}(dyn cm$^{-2}$)  \\
                   \colhead{(1)} &    \colhead{(2)}                &   \colhead{(3)}           & 
               \colhead{(4)}     &    \colhead{(5)}                &    \colhead{(6)}          & 
               \colhead{(7)}     &    \colhead{(8)}                }

\startdata
      knot$-$A/B   &  0.70&    $\ge$    5E15&   $\le$     5E2&   $\ge$     4E06&   1.3E-04&   3E54&   8E-10  \\
      Knot$-$C/G   &  0.73&    $\ge$    1.4E15&   $\le$   8E2&   $\ge$     2E06&   1.2E-04&   2.5E54&   6E-10  \\
      Np$-$1       &  0.85&    $\sim$   1.6E15&   $\sim$  9E2&   $\sim$    2E06&   1.0E-04&   1.6E54&   4E-10  \\
      Sf$-$1       &  0.90&    $\sim$   6E14&   $\sim$    1.4E3&   $\sim$    1.3E06&   1.1E-04&   2E54&   5E-10  \\
      Sf$-\theta$  &  0.90&    $\ge$    2E15&   $\le$     7E2&   $\ge$     2E06&   1.1E-04&   2E54&   5E-10  \\
      Sf$-\eta$    &  0.88&    $\sim$   6E14&   $\sim$    1.5E3&   $\sim$    1.4E06&   1.0E-04&   1.6E54&   4E-10  \\
\enddata
\end{deluxetable}

\clearpage

\clearpage

\clearpage

\clearpage

\clearpage

\clearpage

\clearpage

\end{document}